\documentclass[letter, conference]{ieeeconf}
\IEEEoverridecommandlockouts
\usepackage{cite}
\usepackage{amsmath,amssymb,amsfonts}
\usepackage{standalone}
\usepackage{float}

\usepackage{algorithm}
\usepackage{algorithmic}



\usepackage{varwidth, tikz}
\usetikzlibrary{shapes, arrows, shapes.misc, arrows.meta, positioning, matrix, calc, fit, fadings,patterns, decorations.markings}
\usepackage{relsize}

\newlength{\lw}
\newlength{\lwt}
\newlength{\lnshift}
\newlength{\sshift}
\newlength{\ssshift}
\newlength{\lhalf}
\newlength{\lone}
\newlength{\lonea}
\newlength{\ltwo}
\newlength{\lthree}
\newlength{\vone}

\newlength{\spaceimcstandard}
\newlength{\imcvspace}
\newlength{\imchspace}

\tikzset{
  pics/carc/.style args={#1:#2:#3}{
    code={
      \draw[pic actions] (#1:#3) arc(#1:#2:#3);
    }
  }
}
\def\circledarrowc{
  \draw pic[black,thick]{carc=45:190:0.4em};
  \draw pic[black,thick]{carc=225:370:0.4em};
  \node[draw,fill,single arrow,
    single arrow tip angle=45,
    single arrow head extend=0.75pt,
    single arrow head indent=0pt,
    inner sep=0pt,
    shape border rotate=90] at (0.38em,0.06em) {};
  \node[draw,fill,single arrow,
    single arrow tip angle=45,
    single arrow head extend=0.75pt,
    single arrow head indent=0pt,
    inner sep=0pt,
    shape border rotate=270] at (-0.38em,-0.06em) {};
}

\usepackage{pgfplots}
\pgfplotsset{compat=newest}
\pgfplotsset{
  layers/axis lines on top/.define layer set={
    axis background,
    axis grid,
    axis ticks,
    axis tick labels,
    pre main,
    main,
    axis lines,
    axis descriptions,
    axis foreground,
  }{/pgfplots/layers/standard},
}

\definecolor{colordarkgray}{RGB}{85,85,85}
\definecolor{colorlightgray}{RGB}{179,179,179}
\definecolor{giallo}{RGB}{255,153,0}
\definecolor{blu}{RGB}{102,140,217}
\definecolor{verde}{RGB}{16,150,24}
\definecolor{viola}{RGB}{153,0,153}

\makeatletter

\tikzstyle{chart}=[
    legend label/.style={font={\scriptsize},anchor=west,align=left},
    legend box/.style={rectangle, draw, minimum size=5pt},
    axis/.style={black,semithick,->},
    axis label/.style={anchor=east,font={\tiny}},
]

\tikzstyle{bar chart}=[
    chart,
    bar width/.code={
        \pgfmathparse{##1/2}
        \global\let\bar@w\pgfmathresult
    },
    bar/.style={very thick, draw=white},
    bar label/.style={font={\bf\small},anchor=north},
    bar value/.style={font={\footnotesize}},
    bar width=.75,
]

\tikzstyle{pie chart}=[
    chart,
    slice/.style={line cap=round, line join=round, thin,draw=black},
    pie title/.style={font={\normalfont},align=center},
    slice type/.style 2 args={
        ##1/.style={fill=##2},
        values of ##1/.style={}
    },
    slice type pattern/.style 2 args={
        ##1/.style={pattern=##2},
        values of ##1/.style={}
    }
]

\pgfdeclarelayer{background}
\pgfdeclarelayer{foreground}
\pgfsetlayers{background,main,foreground}

\newcommand{\mypie}[3][]{
    \begin{scope}[#1]
    \pgfmathsetmacro{\curA}{90}
    \pgfmathsetmacro{\r}{1}
    \def\c{(0,0)}
    \node[pie title] at (90:1.3) {#2};
    \foreach \v/\s in{#3}{
        \pgfmathsetmacro{\deltaA}{\v/100*360}
        \pgfmathsetmacro{\nextA}{\curA + \deltaA}
        \pgfmathsetmacro{\midA}{(\curA+\nextA)/2}

        \path[slice,\s] \c
            -- +(\curA:\r)
            arc (\curA:\nextA:\r)
            -- cycle;
        \pgfmathsetmacro{\d}{max((\deltaA * -(.5/50) + 1) , .5)}

        \begin{pgfonlayer}{foreground}
        \ifnum\v>1
        	\path \c -- node[pos=\d,pie values,values of \s,fill=white,inner sep=1pt,font={\scriptsize}]{$\v\%$} +(\midA:\r);
        \fi
        \end{pgfonlayer}

        \global\let\curA\nextA
    }
    \end{scope}
}

\newcommand{\legend}[2][]{
    \begin{scope}[#1]
    \path
        \foreach \n/\s in {#2}
            {
                  ++(0,-10pt) node[\s,legend box] {} +(5pt,-1pt) node[legend label] {\normalsize{\n}}
            }
    ;
    \end{scope}
}

\definecolor{oxforduniversityblue}{RGB}{0,33,71}
\definecolor{oxforduniversityred}{RGB}{204, 41, 0}
\definecolor{colororange}{RGB}{204, 41, 0}  
\definecolor{coloroorange}{HTML}{E65100} 	
\definecolor{colordgray}{HTML}{795548} 		
\definecolor{colorhgray}{HTML}{212121} 		
\definecolor{colorgreen}{RGB}{84, 157, 62} 		
\definecolor{colorlgray}{HTML}{FAFAFA} 		
\definecolor{colorblue}{RGB}{0,33,71}  		
\definecolor{colorred}{RGB}{204, 41, 0}  	
\definecolor{colorpurple}{RGB}{64, 26, 76}	

\usepackage{parskip} 
\begingroup
    \makeatletter
    \@for\theoremstyle:=definition,remark,plain\do{%
        \expandafter\g@addto@macro\csname th@\theoremstyle\endcsname{%
            \addtolength\thm@preskip\parskip
            }%
        }
\endgroup

\usepackage[absolute]{textpos}
\usepackage{balance}
\usepackage{graphicx}
\usepackage{overpic}
\usepackage{rotating}
\usepackage{mathtools}
\usepackage{subcaption}
\usepackage{mymath}
\usepackage{nth}

\makeatletter
\renewcommand*\env@matrix[1][\arraystretch]{%
  \edef\arraystretch{#1}%
  \hskip -\arraycolsep
  \let\@ifnextchar\new@ifnextchar
  \array{*\c@MaxMatrixCols c}}
\makeatother

\usepackage{graphicx}
\usepackage{textcomp}
\usepackage{xcolor}
\def\BibTeX{{\rm B\kern-.05em{\sc i\kern-.025em b}\kern-.08em
    T\kern-.1667em\lower.7ex\hbox{E}\kern-.125emX}}

\usepackage{caption}
\usepackage[hidelinks]{hyperref}
\begin{document}

\setlength{\textfloatsep}{1.5em}
\setlength{\intextsep}{1.5em}

\title{\LARGE \bf Model Predictive Control for Electron Beam Stabilization in a Synchrotron}

\author{Idris Kempf${^*}$, Paul J.\ Goulart${^*}$, Stephen R. Duncan${^*}$ and Michael Abbott${^{**}}$
\thanks{$^*$Corresponding author: {\tt\footnotesize{idris.kempf@eng.ox.ac.uk}}. The authors are with the Department of Engineering Science, University of Oxford, Oxford, UK. This research is supported by the Engineering and Physical Sciences Research Council (EPSRC) with a Diamond CASE studentship.}
\thanks{${^{**}}$Diamond Light Source, Didcot, UK.}
}

\maketitle

\begin{abstract}
Electron beam stabilization in a synchrotron is a disturbance rejection problem, with hundreds of inputs and outputs, that is sampled at frequencies higher than $10$~kHz. In this feasibility study, we focus on the practical issues of an efficient implementation of model predictive control (MPC) for the heavily ill-conditioned plant of the electron beam stabilization problem. To obtain a tractable control problem that can be solved using only a few iterations of the fast gradient method, we investigate different methods for preconditioning the resulting optimization problem and relate our findings to standard regularization techniques from cross-directional control. We summarize the single- and multi-core implementations of our control algorithm on a digital signal processor (DSP), and show that MPC can be executed at the rate required for synchrotron control. MPC overcomes various problems of standard electron beam stabilization techniques, and the successful implementation can increase the stability of photon beams in synchrotron light sources.
\end{abstract}
\section{Introduction}\label{sec:introduction}
A synchrotron light source is a special type of particle accelerator in which charged particles, typically electrons, travel around a circular path called the \emph{storage ring}. When the electrons' paths are bent around the storage ring at relativistic speeds, they lose kinetic energy and emit it in the form of exceptionally bright light, which is used for microscopic experiments. An assembly of magnets produces a magnetic field that confines the electrons in the storage ring. Large magnets steer and focus the electron beam whilst smaller \emph{corrector magnets} attenuate vibrations induced by disturbances and reduce the trajectory error of the electrons down to a few $\mu$m. These disturbances are caused by internal devices, such as the beam light extraction devices, or transmitted through the girders on which the magnet arrays are attached. The position of the electron beam is measured using \emph{beam position monitors} (BPMs) and the corrector magnets are controlled in a feedback loop that is sampled within a frequency range of $10-100$~kHz. The beam trajectory error must be minimized in order to produce high brilliance synchrotron light. This control system is referred to as \emph{fast orbit feedback} and typically has a few hundred BPMs (outputs) and few hundred corrector magnets (inputs).

Diamond Light Source (DLS) is the UK's national synchrotron facility, and its $560$~m circumference storage ring accommodates over $20$ experimental stations. DLS has completed the conceptual design phase of a significant upgrade (DLS-II), which will increase the brightness of the synchrotron light by raising the electron beam energy from $3$~GeV to $3.5$~GeV~\cite{DIAMONDII} and the number of sensors and actuators from $172$ to $252$ and $173$ to $396$, respectively. In the current facility only one type of corrector magnet is used, but DLS-II will instead use separate types for high and low bandwidth correction. In addition, the sampling frequency will be increased from $10$~kHz to $100$~kHz.

The consequences of introducing two types of corrector magnets are twofold. First, the widely used \emph{modal decomposition}~\cite{HEATH} that diagonalizes the input-output transfer function matrix using a singular value decomposition can no longer be applied. Second, amplitude and slew-rate actuator constraints must be considered.

Model predictive control (MPC) allows for an arbitrary number of actuator arrays and provides a systematic way to handle actuator constraints while achieving the same or better disturbance attenuation~\cite{BCINVADMMCONF} as linear control methods. However, the MPC algorithm uses real-time optimization and considerably increases the computational complexity of the fast orbit feedback system. Formulating MPC for the electron beam stabilization problem results in a constrained quadratic program with hundreds of decision variables, and the highly ill-conditioned plant negatively affects the convergence properties of the solver. A tailored MPC implementation is therefore required to obtain an MPC scheme that operates at frequencies higher than $10$~kHz. In anticipation of the upcoming DLS-II upgrade, it was decided to assess the feasibility and performance of installing MPC on the existing DLS-I storage ring. This paper describes an assessment of the design and conception of the future DLS-II fast orbit feedback architecture and allows for an optimal dimensioning of the required controller hardware.

The paper is organized as follows. The process model is introduced in section Section~\ref{sec:preliminaries} and a state-space model and observer introduced in Section~\ref{sec:modelandobserver}. We use standard modelling techniques for setpoint tracking and observer design, but include these details for the benefit of practitioners in the synchrotron community who may be unfamiliar with these methods. We formulate our MPC problem in Section~\ref{sec:mpc}, which we solve using the fast gradient method, and analyze the solver convergence with respect to preconditioning. Finally, Section~\ref{sec:implementation} details the parallel implementation of MPC on a multicore digital signal processor (DSP). The developments presented in this paper apply to DLS-I and II, but the implementation has been tailored to DLS-I.
\section{Preliminaries}\label{sec:preliminaries}
\subsection{Process Model}\label{sec:pm}
For DLS-II, the relationship between the $n_y\!=\!252$ beam displacements $\inRv{\mathbf{y}_k}{n_y}$ measured around the ring, the $n_s\!=\!252$ slow corrector magnets inputs $\inRv{\mathbf{u}_{s,k}}{n_s}$  and the $n_f\!=\!144$ fast corrector magnets inputs $\inRv{\mathbf{u}_{f,k}}{n_f}$ at time $t=k \Delta t$ is given by
\begin{align}
\mathbf{y}_k = \mathbf{R}_s g_s(\inv{z})\mathbf{u}_{s,k} + \mathbf{R}_f g_f(\inv{z})\mathbf{u}_{f,k} + \mathbf{d}_{k},\label{eq:freqdommodel}
\end{align}
where $\Delta t=10$~$\mu$s is the sampling time, $\inv{z}$ represents the backward shift operator and $\mathbf{d}_k$ the disturbances. The matrix $\mathbf{R}\eqdef\left[\mathbf{R}_s\,\mathbf{R}_f\right]\in\R^{n_y\times n_u}$ with $n_u=n_s+n_f$ is called the \emph{orbit response matrix} and typically has a condition number on the order of $10^4$. The scalar transfer functions $g_{(\cdot)}$ model the corrector magnet dynamics plus a transport delay that accounts for unmodeled elements between the central computing node and the power supply of the magnets, and take the form
\begin{align}\label{eq:actuatormodel}
g_{(\cdot)}(\inv{z})= z^{\sm(\mu+1)} \frac{1-e^{\sm a_{(\cdot)}{\Delta t}}}{1-\inv{z}e^{\sm a_{(\cdot)}{\Delta t}}},
\end{align}
where $\mu=10$ is the delay in terms of time steps. The slow magnets have a small bandwidth $a_s = 2\pi \times 100$~Hz but strong a magnetic field, while the fast magnets have a high bandwidth $a_f = 2\pi \times 10$~kHz but a weak magnetic field.

In contrast, the DLS-I storage ring has $n_y=172$ position measurements and $n_u=173$ corrector magnets. Most of the corrector magnets have a medium bandwidth $a_m=2\pi \times700$~Hz, but $n_s=3$ slow and $n_f=2$ fast magnets have been installed for testing purposes in anticipation of the DLS-II upgrade. The DLS-I feedback is sampled at $\Delta t=100$~$\mu$s with a delay of $\mu=7$ time steps.

Note that the vector $\mathbf{y}_k$ describes the displacement in either horizontal or vertical direction perpendicular to the motion of the electron beam. These directions are independent and the electron beam stabilization problem includes two different systems of the form of~\eqref{eq:freqdommodel}. In the following, we will focus on the vertical direction, which is more difficult to control.
\subsection{Cross-Directional Control}\label{sec:cd}
The plant model~\eqref{eq:freqdommodel} is usually referred to as a \emph{cross-directional} system, and similar models are obtained for web forming processes~\cite{HEATH} such as those encountered in paper manufacturing or plastic film extrusion. For cross-directional systems, the response can be split into a spatial component ($\mathbf{R}_{(\cdot)}$) and a temporal component ($g_{(\cdot)}(\inv{z})$). The design of feedback systems for electron beam stabilization has many parallels to cross-directional control~\cite{SANDIRAMULTIARRAY}. In the case of only one type of actuator, a standard approach is to decompose the orbit response matrix using a \emph{singular value decomposition} (SVD) as $\mathbf{R}=\mathbf{U}\mathbf{\Sigma}\trans{\mathbf{V}}$, where $\mathbf{\Sigma}$ may contain blocks of zeros depending on the shape of $\mathbf{R}$,. By defining the \emph{modal} outputs and inputs as $\mathbf{\hat{y}}_k=\trans{\mathbf{U}}\mathbf{y}_k$ and $\mathbf{\hat{u}}_k=\trans{\mathbf{V}}\mathbf{u}_k$, the multi-input multi-output system is decoupled into a set of single-input single-output (SISO) systems. The control input can then be calculated as $\mathbf{u}_k=-\mathbf{V}\mathbf{\hat{K}}c(\inv{z})\mathbf{U}\mathbf{y}_k$, where the gain matrix $\mathbf{\hat{K}}\eqdef \invbr{\trans{\mathbf{\Sigma}}\mathbf{\Sigma}+\lambda I}\trans{\mathbf{\Sigma}}$ with $\lambda>0$ is diagonal and $c(\inv{z})$ is often chosen to be identical for each mode and given by a Dahlin~\cite{SANDIRAWINDUP} or PID~\cite{SCHWARTZ} controller. Regularizing the inverse of $\mathbf{\Sigma}$ is essential to prevent large control gains in the direction of small singular values.

When the system has more than one actuator array, such as in~\eqref{eq:freqdommodel}, the modal decomposition can no longer be applied because the SVDs $\mathbf{R}_{(\cdot)}=\mathbf{U}_{(\cdot)}\mathbf{\Sigma}_{(\cdot)} \trans{\mathbf{V}}_{(\cdot)}$ with ${(\cdot)}=\lbrace\text{s,f}\rbrace$ do \emph{not} share the same matrix of left singular vectors $\mathbf{U}_{(\cdot)}$. In this case, the orbit response matrices can be simultaneously decomposed using alternative methods~\cite{SANDIRAMULTIARRAY,MULTIARRAYGSVD}. Other approaches introduce a frequency deadband between slow and fast actuators and setup two independent control loops~\cite{SCHWARTZ}, which is a suboptimal approach because it prevents control action in the frequency deadband. To handle actuator constraints, the standard controllers must be extended with an anti-windup scheme.
\subsection{Symmetries}\label{sec:sym}
In most synchrotrons the monitors and magnets are placed in repeated patterns around the storage ring. These patterns produce a \emph{circulant} and \emph{centrosymmetric} structure in $\mathbf{R}$~\cite{SYNCSYM}. In contrast to the modal decomposition, which requires the outputs and inputs to be multiplied by dense matrices, the symmetric transformations can be carried out using the computationally efficient Fast Fourier Transformation (FFT). In our previous work, we have shown how these symmetries can be exploited for cross-directional~\cite{SYNCSYM} and MPC~\cite{BCINVADMMCONF} to increase the computational speed of the controller and reduce the memory requirements. These symmetries stand out in the DLS-II orbit response matrix, but have been corrupted in the current orbit response matrix after adding modifications in anticipation DLS-II. Because our MPC algorithm will be tested on the DLS-I storage ring, the structural symmetries are not considered further. However, our implementation could be extended to consider symmetries and would produce significant performance improvements.
\section{Model and Observer}\label{sec:modelandobserver}
\subsection{State-Space System}\label{sec:ss}
The standard linear MPC formulation requires a state-space model, and we choose to define the states $\mathbf{x}_{(\cdot),k}\in\R^{n_{(\cdot)}}$ as
\begin{align}\label{eq:statechoice}
\mathbf{x}_{(\cdot),k} = z^{-1}\frac{1-e^{\sm a_{(\cdot)}{\Delta t}}}{1-\inv{z}e^{\sm a_{(\cdot)}{\Delta t}}} \mathbf{u}_{(\cdot),k},
\end{align}
where ${(\cdot)}=\lbrace\text{s,f}\rbrace$. Applying the backward shift operator to $\mathbf{x}_{(\cdot),k}$ and $\mathbf{u}_{(\cdot),k}$ yields a state-space representation of~\eqref{eq:freqdommodel} as
\begin{equation}\begin{aligned}\label{eq:sssystem}
\begin{pmatrix}\mathbf{x}_{s,k+1}\\\mathbf{x}_{f,k+1}\end{pmatrix}
&\!=\!
\begingroup
\setlength\arraycolsep{2pt}
\begin{bmatrix}\mathbf{A}_s & 0\\ 0 & \mathbf{A}_f\end{bmatrix}
\endgroup
\!\begin{pmatrix}\mathbf{x}_{s,k}\\\mathbf{x}_{f,k}\end{pmatrix}
\!+\!
\begingroup
\setlength\arraycolsep{2pt}
\begin{bmatrix}\mathbf{B}_s & 0\\ 0 & \mathbf{B}_f\end{bmatrix}
\endgroup
\!\begin{pmatrix}\mathbf{u}_{s,k}\\\mathbf{u}_{f,k}\end{pmatrix},\\
\mathbf{y}_k &\!=\!
\begingroup
\setlength\arraycolsep{3pt}
\begin{bmatrix}\mathbf{R}_s & \mathbf{R}_f\end{bmatrix}
\endgroup
\!\begin{pmatrix}\mathbf{x}_{s,k\sm\mu}\\\mathbf{x}_{f,k\sm\mu}\end{pmatrix}+\mathbf{d}_k,
\end{aligned}\end{equation}
where $\mathbf{A}_{(\cdot)} = I e^{\sm a_{(\cdot)}{\Delta t}}$ and $\mathbf{B}_{(\cdot)} = I -\mathbf{A}_{(\cdot)}$. In the form~\eqref{eq:sssystem}, the states $\mathbf{x}_{s,k}$ and $\mathbf{x}_{f,k}$ are proportional to the magnetic fields of the slow and fast correctors acting on the electron beam. We will use the more compact notation
\begin{equation}\begin{aligned}\label{eq:ssshort}
\mathbf{x}_{k+1}=\mathbf{A}\mathbf{x}_{k}+\mathbf{B}\mathbf{u}_{k},\qquad
\mathbf{y}_{k}=\mathbf{C}\mathbf{x}_{k-\mu}+\mathbf{d}_{k},
\end{aligned}\end{equation}
where $\mathbf{x}_{k}\!\!\eqdef\!\!(\mathbf{x}_{s,k}^\Tr,\mathbf{x}_{f,k}^\Tr)^\Tr$ and $\mathbf{u}_{k}\!\!\eqdef\!\!(\mathbf{u}_{s,k}^\Tr,\mathbf{u}_{f,k}^\Tr)^\Tr$. A widely used control approach for~\eqref{eq:ssshort} is the \emph{linear quadratic regulator} (LQR) that computes a control law as $\mathbf{u}_k=-\mathbf{K}\mathbf{x}_k$ and can be interpreted as an unconstrained version of MPC.

In practice, the actuator inputs (currents) $\mathbf{u}_{s,k}$ and $\mathbf{u}_{f,k}$ are subjected to slew-rate constraints and amplitude constraints, respectively. The constraints can be modeled as
\begin{subequations}\label{eq:constraints}
\begin{align}
\mathcal{U}_\text{a} &= \set{\mathbf{u}_{f,k}\in\R^{n_f}}{\sm \alpha \leq \mathbf{u}_{f,k} \leq \alpha},\label{eq:ampl}\\
\mathcal{U}_\text{r} &= \set{\mathbf{u}_{s,k},\mathbf{u}_{s,k\sm 1}\in\R^{n_s}\!}{\!\sm \rho \leq \mathbf{u}_{s,k}\sm \mathbf{u}_{s,k\sm 1} \leq \rho},\label{eq:rate}
\end{align}
\end{subequations}
where the inequalities are to be read component-wise. The magnitude of the amplitude limit $\alpha$ depends on the normalization of the inputs and the slew-rate constant $\rho$ is chosen as $\rho=\alpha/10$, which reflects results obtained from preliminary simulations of the fast corrector magnets. We consider symmetric limits on both slew-rate and amplitude, but the algorithm is easily modified to allow asymmetric limits. Analogous to the shorthand notation~\eqref{eq:ssshort}, we will abbreviate~\eqref{eq:constraints} as $\mathbf{u}_k\in\mathcal{U}$. Note that in our implementation we will assume that slow \emph{and} fast actuators are constrained by both slew-rate and amplitude constraints, but the limits for each actuator type are adjusted accordingly.
\subsection{Setpoint Calculation}\label{sec:setp}

The aim of the control system is to reject the disturbances $\mathbf{d}_k$ in~\eqref{eq:freqdommodel}. In response to a constant disturbance, a zero steady-state output $\mathbf{y}_k$ requires the open-loop transfer function of~\eqref{eq:freqdommodel} to have integrating behavior~\cite{SKOGESTADMULTI}. Because the plant transfer functions $g_{(\cdot)}(\inv{z})$ lack integrating behavior, the controller must implement the integrator. For an LQR approach, there exist several methods to add integrating behavior. One way is to augment the system with a set of output integrators. However, this method would slow down the subsequent MPC algorithm by increasing the number of optimization variables. Alternatively, one can compute the setpoints $\mathbf{\bar{u}}$ and $\mathbf{\bar{x}}$ and use the feedback law $\mathbf{u}_k=\mathbf{\bar{u}}+\mathbf{u}_k^\star$~\cite{DISTMPC}, where $\mathbf{u}_k^\star$ is obtained from $\mathbf{u}_k^\star=\sm\mathbf{K}\mathbf{x}_k$ in the case of LQR or as the solution to an optimization problem in the case of MPC. The setpoints should be calculated such that $\lim_{k\rightarrow\infty}\mathbf{y}_k=0$, which using~\eqref{eq:ssshort} yields
\begin{align}\label{eq:linsyssp}
\begin{pmatrix}0\\ \mathbf{\bar{d}}_k\end{pmatrix}=
\begin{bmatrix}I-\mathbf{A} & -\mathbf{B}\\-\mathbf{C} & 0\end{bmatrix}
\begin{pmatrix}\mathbf{\bar{x}}_k\\\mathbf{\bar{u}}_k\end{pmatrix}
\reqdef\mathbf{S}\begin{pmatrix}\mathbf{\bar{x}}_k\\\mathbf{\bar{u}}_k\end{pmatrix},
\end{align}
where $\inR{\mathbf{S}}{n_u+n_y}{2n_u}$ and $\mathbf{\bar{d}}_k\in\R^{n_y}$ is a disturbance estimate that is obtained from the observer. The coefficient matrix $\inR{\mathbf{S}}{n_u+n_y}{2n_u}$ has more columns than rows and the Moore-Penrose pseudoinverse $\pinv{\mathbf{S}}=\invbr{\trans{\mathbf{S}}\mathbf{S}}\trans{\mathbf{S}}$ can be used to solve for $\mathbf{\bar{x}}_k$ and $\mathbf{\bar{u}}_k$. Note the zeros in the left-hand side vector of~\eqref{eq:linsyssp}, so that in practice, only the last $n_y$ columns of $\pinv{\mathbf{S}}$ need to be considered.
\subsection{State and Disturbance Observer}\label{sec:obs}
Standard methods from cross-directional control use output feedback to control~\eqref{eq:freqdommodel}, whereas LQR and MPC use state feedback to control the equivalent state-space system~\eqref{eq:ssshort}. The states $\mathbf{x}_k$ and disturbances $\mathbf{d}_k$ are not measurable and these values must be inferred from the measured outputs using an observer. The observer continuously computes the state-transition equation in~\eqref{eq:ssshort} and adds the term $\mathbf{L}(\mathbf{y}_k-\mathbf{C}\mathbf{x}_k)$, where we chose the observer gain $\mathbf{L}$ as the steady-state Kalman filter gain~\cite{OPTSTATEESTIM}.

For modelling the disturbance, a first-order model that is driven by zero-mean independent and identically distributed white noise~\cite{DISTMPC} is used, i.e.
\begin{align}\label{eq:distmodel}
\mathbf{d}_{k+1} = \mathbf{A}_d \mathbf{d}_k +\mathbf{v}_k,
\end{align}
where $\mathbf{v}_k\sim\mathcal{N}(0,\sigma_\mathbf{v}^2)$ and we choose $\mathbf{A}_d=I$. Alternatively, the matrix $\mathbf{A}_d$ can be obtained from a first-order autoregressive fit from the measurement data.

All measurements of system~\eqref{eq:ssshort} are delayed by $\mu$ time steps and the incoming measurement $\mathbf{y}_k$ at time $t=k\Delta t$ contains information about the state $\mathbf{x}_{k\sm\mu}$ at time $t=(k-\mu)\Delta t$. One possibility to integrate the delayed measurements is to formulate a delay-free system by augmenting~\eqref{eq:ssshort} with $\mu \times (n_s+n_f)$ states, i.e. defining $\mathbf{z}_k^i\eqdef \mathbf{x}_{k-i}$, $i=1,\dots,7$ and adding $\mathbf{z}_{k+1}^{1}= \mathbf{x}_{k}$ and $\mathbf{z}_{k+1}^{i+1}= \mathbf{z}_{k}^i$ and rewriting the state transition equations as
\begin{equation}\label{eq:obs}\begin{aligned}
\begin{pmatrix}\mathbf{\hat{x}}_{k+1}\\
\mathbf{\hat{z}}_{k+1}^1\\
\vdots\\
\mathbf{\hat{z}}_{k+1}^\mu\\
\mathbf{\hat{d}}_{k+1}\end{pmatrix}=
&\begin{bmatrix}
\mathbf{A} &  &  & \\
I & 0 &  & \\[-4pt]
0 & \ddots & \ddots & \\[-4pt]
 & \ddots & I & 0\\
 & & 0& \mathbf{A}_d
\end{bmatrix}
\begin{pmatrix}
\mathbf{\hat{x}}_{k}\\
\mathbf{\hat{z}}_{k}^1\\
\vdots\\
\mathbf{\hat{z}}_{k}^\mu\\
\mathbf{\hat{d}}_{k}\end{pmatrix}+
\begin{bmatrix}\mathbf{B}\\0\\\vdots\\0\end{bmatrix}\mathbf{u}_k\\&+\mathbf{L}\left(\mathbf{y}_k-\mathbf{C}\mathbf{\hat{z}}_{k}^\mu-\mathbf{\hat{d}}_{k}\right),
\end{aligned}\end{equation}
where variables with a hat denote estimated quantities and the state-space system~\eqref{eq:ssshort} has been combined with the disturbance model~\eqref{eq:distmodel}.

The observer~\eqref{eq:obs} requires a matrix-vector multiplication with a dense $\inR{\mathbf{L}}{((\mu+1)n_u+n_y)}{n_y}$, which is a computationally expensive operation that can be avoided as follows. First, partition the observer gain as $\mathbf{L} = [\mathbf{L}_\mathbf{x}^\Tr,\,\mathbf{L}_{\mathbf{z}^1}^\Tr,\dots\,\mathbf{L}_{\mathbf{z}^\mu}^\Tr,\,\mathbf{L}_{\mathbf{d}}^\Tr]^\Tr$, where the partitioning of $\mathbf{L}$ matches the partitioning of the vector on the left-hand side of~\eqref{eq:obs}. Then, update the most delayed state $\mathbf{\hat{z}}^\mu$ and the disturbance estimate $\mathbf{\hat{d}}$ using $\mathbf{L}_{\mathbf{z}^\mu}$ and $\mathbf{L}_{\mathbf{d}}$, respectively, and reserve $\Delta\mathbf{\hat{y}}_k\eqdef \mathbf{L}_{\mathbf{z}^\mu}(\mathbf{y}_k-\mathbf{C}\mathbf{\hat{z}}_{k}^\mu-\mathbf{\hat{d}}_{k})$. Finally, update the states $\mathbf{\hat{z}}^i$, $i=1,\dots,\mu\sm 1$ by adding $\mathbf{A}^{\mu\sm i}\Delta\mathbf{\hat{y}}_k$ and in particular $\mathbf{\hat{x}}$ using $\mathbf{A}^{\mu}\Delta\mathbf{\hat{y}}_k$. Note that the matrices $\mathbf{A}^i$ are diagonal and can be pre-computed offline.
\section{Model Predictive Control}\label{sec:mpc}
\subsection{Problem Formulation}\label{sec:mpcformulation}
At time $t=k\Delta t$, the MPC scheme computes a control input by predicting the future evolution of the system and minimizing a quadratic objective function over the planning horizon $N$, while considering inputs that lie in the constraint set~\eqref{eq:constraints} only. This can be achieved via repeated solution of the following constrained quadratic program (CQP):
\begin{gather}\label{eq:mpc}
\begin{aligned}
\min &\sum_{i=0}^{N\sm 1} \xnorm{\mathbf{x}_i-\mathbf{\bar{x}}}{\mathbf{Q}}^2+\xnorm{\mathbf{u}_i-\mathbf{\bar{u}}}{\mathbf{R}}^2 + \xnorm{\mathbf{x}_N-\mathbf{\bar{x}}}{\mathbf{P}}^2\\
\text{s.t.}\,&\qquad\mathbf{x}_{i+1} = \mathbf{A}\mathbf{x}_i+\mathbf{B}\mathbf{u}_i,\,\,\mathbf{x}_0=\mathbf{\hat{x}}_k,\,\,\mathbf{u}_i\in\mathcal{U},\\
     &\qquad\mathbf{y}_i = \mathbf{C}\mathbf{x}_i,\\
\end{aligned}
\end{gather}
for $i=0,\dots,N\sm 1$, where the optimization variables are $\mathbf{x}_{(\cdot)}$ and $\mathbf{u}_{(\cdot)}$. Even though the solution of~\eqref{eq:mpc} is a sequence of inputs $\mathbf{u}_0^\star,\dots,\mathbf{u}_{N\sm 1}^\star$, only the first input $\mathbf{u}_0^\star$ is applied to the plant and the optimization repeated at the next time step $t+\Delta t$. The matrices $\mathbf{Q}\eqdef\mathbf{C}^\Tr\mathbf{C}$ and $\mathbf{R}$ are the state and output weighting matrices, respectively, while $\mathbf{P}=\trans{\mathbf{P}}\succ 0$ is the terminal cost matrix. The optimization problem~\eqref{eq:mpc} has a unique solution if $\mathbf{R}\succ 0$, $\mathbf{Q}\succeq 0$ and if the pairs $(\mathbf{A},\mathbf{B})$ and $(\mathbf{A},\mathbf{Q}^{\frac{1}{2}})$ are controllable and observable, respectively~\cite[Ch. 12]{MPCBOOK}. Because the system~\eqref{eq:ssshort} is stable and there are no state constraints, the MPC scheme is guaranteed to be feedback stable if the terminal cost matrix $\mathbf{P}$ is obtained from the \emph{discrete-time Riccati equation} (DARE) associated with the unconstrained LQR,
\begin{align}\label{eq:dare}
\mathbf{A}^\Tr \mathbf{P} \mathbf{A} - \mathbf{A}^\Tr \mathbf{P} \mathbf{B} \left(\mathbf{B}^\Tr \mathbf{P} \mathbf{B} +\mathbf{R}\right)^{-1}\mathbf{B}^\Tr \mathbf{P} \mathbf{A} + \mathbf{Q} =\mathbf{P},
\end{align}
where we choose the matrices $\mathbf{Q}$ and $\mathbf{R}$ to be the same as in~\eqref{eq:mpc}.

By defining $\mathbf{x}=(\mathbf{x}_0^\Tr,\dots,\mathbf{u}_{N}^\Tr)^\Tr$ and $\mathbf{u}\eqdef(\mathbf{u}_0^\Tr,\dots,\mathbf{u}_{N\sm 1}^\Tr)^\Tr$, the state-transition equations $\mathbf{x}_{i+1} = \mathbf{A}\mathbf{x}_i+\mathbf{B}\mathbf{u}_i$ can be rewritten as $\mathbf{x}=\mathbf{G}\mathbf{u}+\mathbf{H}\mathbf{x}_0$, where
\begin{align*}
\mathbf{G}=
\begin{bmatrix}
0  & \dots \\
\mathbf{B}  &   &   \\
\mathbf{A}\mathbf{B} & \mathbf{B} &   &  \\[-0.5em]
\vdots & & \ddots\\
\mathbf{A}^{N\sm 1}\mathbf{B} & \mathbf{A}^{N\sm2}\mathbf{B} & \dots & \mathbf{B}
\end{bmatrix},\qquad
\mathbf{H}=
\begin{bmatrix}
I \\ \mathbf{A} \\ \mathbf{A}^2\\[-0.5em]\vdots \\ \mathbf{A}^N
\end{bmatrix}.
\end{align*}
By substituting $\mathbf{x}=\mathbf{G}\mathbf{u}+\mathbf{H}\mathbf{x}_0$, the states $\mathbf{x}$ can be eliminated from~\eqref{eq:mpc}, producing the equivalent condensed problem
\begin{align}\label{eq:qpshort}
\min_{\inRv{\mathbf{u}}{Nn_u}} \frac{1}{2}\trans{\mathbf{u}}\mathbf{J}\mathbf{u}+\trans{\mathbf{q}}\mathbf{u}\quad\text{s.t.}\quad\mathbf{u}\in\mathcal{U}_N,
\end{align}
where $\mathcal{U}_N=\mathcal{U}\times\dots\times\mathcal{U}$ and $\mathbf{J}$ and $\mathbf{q}$ are obtained as
\begin{subequations}\begin{align}
\mathbf{J} &\eqdef \mathbf{G}^\Tr \left( (I_N\otimes \mathbf{Q})\oplus \mathbf{P} \right)\mathbf{G}+(I_N\otimes \mathbf{R}),\label{eq:Jdef}\\
\begin{split}
\mathbf{q} &\eqdef
\mathbf{G}^\Tr \left( (I_N\otimes \mathbf{Q})\oplus \mathbf{P} \right)\mathbf{H}\mathbf{x}_0
-\mathbf{G}^\Tr\begin{bmatrix}\mathbf{1}_N\otimes\mathbf{Q}\\\mathbf{P}\end{bmatrix}\mathbf{\bar{x}}\\&\qquad-(\mathbf{1}_N\otimes \mathbf{R})\mathbf{\bar{u}},\label{eq:qdef}
\end{split}
\end{align}\end{subequations}
with $\otimes$ and $\oplus$ denoting the Kronecker product and block-diagonal concatenation, respectively, $I_N$ the identity matrix of size $N\times N$ and $\mathbf{1}_N$ a vector of ones of length $N$. Note that the slew-rate constraints couple the inputs across horizon stages and the set $\mathcal{U}_N$ depends on the previously calculated input $\mathbf{u}_{k\sm 1}^\star$. After finding a solution to~\eqref{eq:qpshort}, the set $\mathcal{U}_N$ must therefore be updated as well as the vector $\mathbf{q}$ on the arrival of a new measurement. In practice, we substitute~\eqref{eq:linsyssp} in~\eqref{eq:qdef} to avoid computing the setpoints $\mathbf{\bar{x}}$ and $\mathbf{\bar{u}}$.
\begin{figure}
\begin{center}
\hspace{1.1cm}\ref{ibmlegend}
\begin{tikzpicture}
\begin{semilogxaxis}[set layers,
    every axis plot/.append style={on layer=pre main},
xlabel={Frequency [Hz]},
ylabel={IBM [$\mu$m]},
legend pos=north west,
name=plot1,
major grid style={line width=.2pt,draw=black},
width=\linewidth,
height=0.6\linewidth,
legend columns = 3,
legend cell align={left},
legend style={/tikz/every even column/.append style={column sep=0.2cm}},
legend to name=ibmlegend,
ymin=0, ymax=2,
xmin=0, xmax=5040,
xmajorgrids,
major grid style={line width=.2pt,draw=gray!50,dotted},
]
\addplot [color=colorlightgray, mark=pentagon, mark size=3] table [x=a, y=f, col sep=comma] {mpc_data.csv};
\addlegendentry{Off}
\addplot [color=black, mark=triangle, mark size=3] table [x=a, y=e, col sep=comma] {mpc_data.csv};
\addlegendentry{IMC}
\addplot [color=black, mark=star, mark size=3] table [x=a, y=g, col sep=comma] {mpc_data.csv};
\addlegendentry{IMC constr.}
\addplot [color=colorred, mark=square, mark size=3] table [x=a, y=b, col sep=comma] {mpc_data.csv};
\addlegendentry{MPC (1)}
\addplot [color=colorblue, mark=diamond, mark size=4] table [x=a, y=c, col sep=comma] {mpc_data.csv};
\addlegendentry{MPC (2)}
\addplot [color=colorgreen, mark=o, mark size=3] table [x=a, y=d, col sep=comma] {mpc_data.csv};
\addlegendentry{MPC (8)}
\end{semilogxaxis}

\begin{axis}[at={(50,80)}, anchor = south west,
axis background/.style={fill=gray!10},
xmin=75,xmax=81,
ymin=0.172,ymax=0.176,
width=0.35\linewidth,
height=0.3\linewidth,
xtick={76,80},
ytick={0.173,0.175},
y tick label style = {font=\tiny,
					  /pgf/number format/.cd,
            			fixed,
            			fixed zerofill,
            			precision=3,
        			  /tikz/.cd},
x tick label style = {font=\tiny}
]
\addplot [color=colorred, mark=square, mark size=2] table [x=a, y=b, col sep=comma] {mpc_data_lf.csv};
\addplot [color=colorblue, mark=diamond, mark size=2] table [x=a, y=c, col sep=comma] {mpc_data_lf.csv};
\addplot [color=colorgreen, mark=o, mark size=2] table [x=a, y=d, col sep=comma] {mpc_data_lf.csv};
\end{axis}
\end{tikzpicture}
\end{center}
\caption{Integrated beam motion (IBM) for the uncontrolled beam, IMC without and with applied constraints and MPC~($N$) with $N$ denoting the horizon.}\label{fig:ibm}
\end{figure}
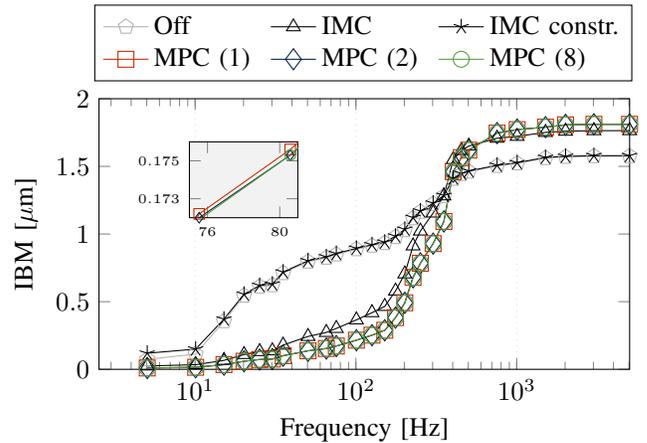
\subsection{Synchrotron Performance Metric}
The performance of the control algorithm can be evaluated using the integrated beam motion (IBM), which is defined as the square root of $\sum_{f=0}^F\frac{2}{F^2}\abs{y_i(f)}^2$, where $y_i(f)$ is the discrete Fourier transform (DFT) of monitor output $i$ and $F$ the frequency in Hz. The IBM is the discrete integral of the DFT of $\mathbf{S}_i(\inv{z})\mathbf{d}_k$, where $\mathbf{S}_i(\inv{z})$ is the sensitivity transfer function matrix of output $i$. Fig.~\ref{fig:ibm} shows the IBM averaged over all monitors for different horizons $N$ and for a particular choice of $\mathbf{Q}$ and $\mathbf{R}$ that will be discussed in Section~\ref{sec:conditioning}. The figure also shows the uncontrolled beam, the simulated IMC for the unconstrained system and IMC for the case that the computed inputs are clipped using~\eqref{eq:constraints}. It can be seen that there is little performance improvement $N$ larger than $1$ or $2$.
Compared to the unconstrained IMC, MPC performs slightly better for lower frequencies but slightly worse for higher frequencies. This ``waterbed'' effect can be controlled by tuning the weighting matrices. Because the computation time is limited to $100$~$\mu$s and we see little improvement for larger horizons, we consider only $N\leq 2$ in the following.
\subsection{Fast Gradient Method}\label{sec:fgm}
Suitable algorithms for solving the CQP~\eqref{eq:qpshort} can be split into first-order methods, such as the \emph{fast gradient method} (FGM) and the \emph{alternating direction method of multipliers} (ADMM), and second-order methods, such as the \emph{interior-point method}. First-order methods use only the first derivative of the objective function, while second-order methods also use the second derivative. First-order methods typically converge quickly to a low-accuracy solution with a low per-iteration computational cost, but need far more iterations to achieve a high-accuracy solution. By contrast, second-order methods need fewer iterations to achieve a high-accuracy solution, but also have a higher per-iteration computational cost. A low-accuracy solution produced by a first-order algorithm is sufficient for an MPC problem~\cite{OSQPMATHPROG}. In~\cite{FGMDYKSTRA}, we showed how FGM outperforms ADMM in terms of computational speed for our particular constraint set. The convergence of ADMM is less affected by ill-conditioned problem data than the FGM, but the algorithm augments the vector $\mathbf{u}$ in~\eqref{eq:qpshort} to accommodate the constraints, which slows down the implementation on the DSP.

The FGM is summarized in Alg.~\ref{alg:fgm} (lines 5-10) with a constant step size $\beta = (\sqrts{\lambda_{max}}-\sqrts{\lambda_{min}})/(\sqrts{\lambda_{max}}+\sqrts{\lambda_{min}})$, where $\lambda_\text{min}$ and $\lambda_\text{max}$ are the minimum and maximum eigenvalues of the Hessian $\mathbf{J}$~\cite[Ch. 2.2]{OPTIMNESTEROV}. In contrast to ADMM, the FGM does not require augmentation of the decision variables but applies the projection operator $\mathcal{P}_{\mathcal{U}_N}$ onto $\mathcal{U}_N$ instead. For $N\!=\!1$, the projection simply limits each component of $\mathbf{u}$ to a minimum and maximum value given by~\eqref{eq:ampl} or~\eqref{eq:rate}. For $N\!=\!2$, the projection onto~\eqref{eq:rate} is more complicated and consists of projecting the pairs $(\mathbf{u}_{0}^i, \mathbf{u}_{1}^i)$, where $i$ denotes the $i$th actuator, onto a hexagon with corner points that depend on the input $\mathbf{u}_{\sm 1}^i$ calculated at time step $k-1$~\cite{FGMDYKSTRA}. Note that on line 5, we warm-start by initializing the FGM using the input calculated at time $(k-1)\Delta t$, which considerably improves the convergence properties of the algorithm~\cite{CERTIFMPC}. Lines marked with the circled arrow \tikz{\circledarrowc} denote synchronization steps of the parallel implementation (Section~\ref{sec:par}) and we will consider the fixed iteration number $I_\text{max}$ in Section~\ref{sec:conditioning}.
\begin{algorithm}[H]
 \caption{MPC for electron beam stabilization}\label{alg:fgm}
 \begin{algorithmic}[1]
    \REQUIRE $\mathbf{y}_k$
    \ENSURE $\mathbf{u}_k$
	\STATE Transfer $\mathbf{y}_k$\hfill \tikz{\circledarrowc}
    \STATE Update observer $\Rightarrow$ $\mathbf{\hat{x}}_k, \mathbf{\hat{d}}_k$ \hfill \tikz{\circledarrowc}\hspace{0.25cm}\tikz{\circledarrowc}
    \STATE Update $\mathbf{q}=\mathbf{q}(\mathbf{\hat{x}}_k, \mathbf{\hat{d}}_k)$\hfill \tikz{\circledarrowc}
	\STATE Update $\mathcal{U}_N=\mathcal{U}_N(\mathbf{u}_{k\sm 1})$
	\STATE Set $\mathbf{v}_i=\mathbf{u}_{k\sm 1}$ and $\mathbf{p}_i=0$
	\FOR{$i = 0$ to $I_{max}$}
  	  \STATE $\mathbf{t}_i = (I - \mathbf{J}\inv{\lambda}_{max})\mathbf{v}_i - \mathbf{q} \inv{\lambda}_{max}$ \hfill \tikz{\circledarrowc}
  	  \STATE $\mathbf{p}_{i+1} = \mathcal{P}_{\mathcal{U}_N}(\mathbf{t}_i)$
  	  \STATE $\mathbf{v}_{i+1} = (1+\beta)\mathbf{p}_{i+1} - \beta \mathbf{p}_i$
	\ENDFOR \hfill \tikz{\circledarrowc}
	\STATE Transfer $\mathbf{u}_k=\mathbf{p}_{I_{max}+1}$
 \end{algorithmic}
\end{algorithm}
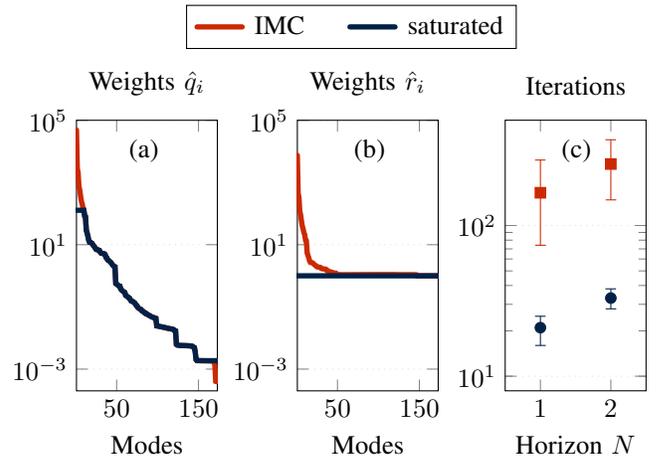
\begin{figure}\begin{center}{
\hspace{2em}\ref{named}\vspace{0.5em}
\begin{tikzpicture}
\begin{semilogyaxis}[xlabel={Modes},
title={Weights $\hat{q}_i$},
name=plot1,
major grid style={line width=.2pt,draw=black},
width=0.4\linewidth,
height=0.6\linewidth,
legend columns = 2,
legend to name=named,
legend entries={$\,$IMC, $\,$saturated},
legend cell align={left},
legend style={/tikz/every even column/.append style={column sep=0.5cm}},
ymin=2*10^(-4), ymax=10^5,
xmin=1, xmax=173,
xtick={50,150},
ymajorgrids,
major grid style={line width=.2pt,draw=gray!50,dotted},
y label style={yshift=-0.9em},
]
\addplot [color=colorred,line width=2pt,line cap=round] table [x=a, y=b, col sep=comma] {weightings.csv};
\addplot [color=colorblue,line width=2pt,line cap=round] table [x=a, y=d, col sep=comma] {weightings.csv};
\end{semilogyaxis}
\begin{semilogyaxis}[xshift=0.34\linewidth,
xlabel={Modes},
xtick={50,150},
title={Weights $\hat{r}_i$},
legend pos=north west,
name=plot2,
major grid style={line width=.2pt,draw=black},
width=0.4\linewidth,
height=0.6\linewidth,
ymin=2*10^(-4), ymax=10^5,
xmin=1, xmax=173,
ymajorgrids,
major grid style={line width=.2pt,draw=gray!50,dotted},
]
\addplot [color=colorred,line width=2pt,line cap=round] table [x=a, y=c, col sep=comma] {weightings.csv};
\addplot [color=colorblue,line width=2pt,line cap=round] table [x=a, y=e, col sep=comma] {weightings.csv};
\end{semilogyaxis}
\begin{semilogyaxis}[xshift=0.66\linewidth,
width=0.4\linewidth,
height=0.6\linewidth,
xlabel=Horizon $N$,
xtick={1,2},
xmin=0.5, xmax=2.5,
ymin=8,ymax=5*10^2,
ytickten={1,...,2},
title={Iterations},
name=plot1,
ymajorgrids,
major grid style={line width=.2pt,draw=gray!50,dotted},
y label style={yshift=-0.4em},
legend columns = 2,
legend to name=namedb,
legend entries={$\,\mathbf{J_\text{sat}}$, $\,\mathbf{J}$},
legend cell align={left},
legend style={/tikz/every even column/.append style={column sep=0.5cm}},]
\addplot [color=colorblue, only marks, mark size=2,mark=*]
 plot [error bars/.cd, y dir = both, y explicit]
 table[x =a, y =b, y error plus=d, y error minus=c, col sep=comma]{iter_trunc_sat.csv};
\addplot [color=colorred, only marks, mark size=2, mark=square*]
 plot [error bars/.cd, y dir = both, y explicit]
 table[x =a, y =e, y error plus=g, y error minus=f, col sep=comma]{iter_trunc_sat.csv};
\end{semilogyaxis}

\node[] at (0.9,3.2) {(a)};
\node[] at (3.9,3.2) {(b)};
\node[] at (6.65,3.2) {(c)};
\end{tikzpicture}}\end{center}
\caption{(a) State and (b) input weights and (c) corresponding average FGM iteration number.}\label{fig:weights}
\end{figure}%
\subsection{Preconditioning of the Hessian}\label{sec:conditioning}
In Alg.~\ref{alg:fgm}, we have chosen a fixed number of iterations $I_\text{max}$ rather than using a stopping criterion, which would increase the computational complexity. This is common in embedded systems applications, and an upper iteration bound can be obtained ~\cite{CERTIFMPC} from
\begin{align}\label{eq:iterbound}
I_\text{max}\!=\!\max\left\lbrace\!0,\,
\min \left\lbrace
\!\ceil*{\!\frac{\ln\epsilon \!\sm\!\ln\Delta}{\ln(1\!\sm\!\sqrt{\frac{1}{\kappa}})}\!},
\ceil*{\!2\sqrt{\frac{\Delta}{\epsilon}}\!\sm\!2}\!
\right\rbrace\!\!\right\rbrace,
\end{align}
where $\epsilon=10^{\sm 3}$ is the desired solution accuracy, $\kappa \eqdef\kappa(\mathbf{J})$ is the condition number of the Hessian and $\Delta$ is a constant that depends on the constraint set $\mathcal{U}_N$. From~\eqref{eq:iterbound}, it can be seen that if $\kappa$ is large, then $I_\text{max}$ tends to be large. For $N\!=\!1$, $\mathbf{Q}=\trans{\mathbf{C}}\mathbf{C}$ and $\mathbf{R}=I$, $\kappa(\mathbf{J})\approx 6000$, which is far too large to solve Alg.~\ref{alg:fgm} at $10$~kHz.
%
The condition number of the Hessian can be reduced by setting $\mathbf{R}=rI$ with $r\gg 1$, but the performance of the controller then rapidly degrades. Alternatively, the Hessian can be preconditioned using an invertible transformation matrix $\inR{\mathbf{E}}{n_u}{n_u}$ such that the condition number of the Hessian $\invbrT{I_N\otimes\mathbf{E}}\mathbf{J}\invbr{I_N\otimes\mathbf{E}}$ is minimized. The matrix $\mathbf{E}$ can be found using semidefinite programming methods~\cite[Ch. 3.1]{LMIBOOK}. However, choosing a dense $\mathbf{E}$ significantly increases the computational complexity of the FGM as the projection is complicated. If $\mathbf{E}$ is instead restricted to be a diagonal matrix, then $\kappa(\mathbf{J})$ is not improved substantially.
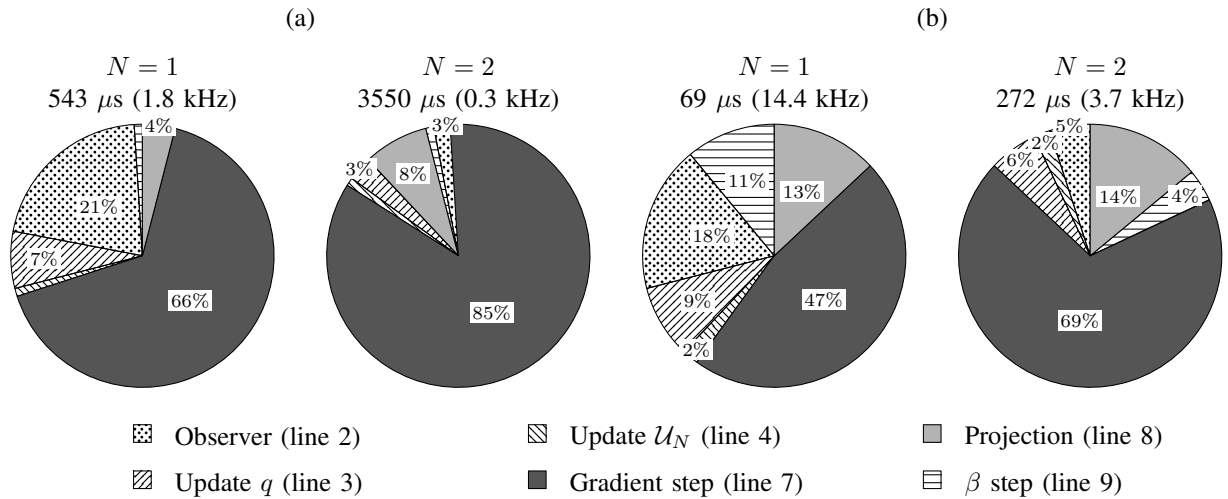
\begin{figure*}
\begin{center}
\begin{tikzpicture}
[
    pie chart,
    slice type={gradstep}{colordarkgray},
    slice type pattern={observer}{crosshatch dots},
    slice type pattern={initq}{north east lines},
    slice type={projection}{colorlightgray},
    slice type pattern={initprojection}{north west lines},
    slice type pattern={betastep}{horizontal lines},
    pie values/.style={font={\small}},
    scale=1.75
]
    \mypie[values of projection/.style={pos=1}]{$N=1$\\543~$\mu$s (1.8~kHz)}{1/betastep,21/observer,7/initq,1/initprojection,66/gradstep,4/projection}
    \mypie[xshift=2.4cm,values of observer/.style={pos=1},values of initq/.style={pos=1}]%
        {$N=2$\\3550~$\mu$s (0.3~kHz)}{3/observer,1/betastep,8/projection,3/initq,1/initprojection,85/gradstep}
    \mypie[xshift=4.8cm]%
        {$N=1$\\69~$\mu$s (14.4~kHz)}{11/betastep,18/observer,9/initq,2/initprojection,47/gradstep,13/projection}
    \mypie[xshift=7.2cm,values of observer/.style={pos=1},values of initq/.style={pos=0.9}]%
        {$N=2$\\272~$\mu$s (3.7~kHz)}{5/observer,2/initprojection,6/initq,69/gradstep,4/betastep,14/projection}
	\legend[shift={(0cm,-1cm)}]{{Observer (line 2)}/observer,{Update $q$ (line 3)}/initq}
	\legend[shift={(3cm,-1cm)}]{{Update $\mathcal{U}_N$ (line 4)}/initprojection,{Gradient step (line 7)}/gradstep}
	\legend[shift={(6cm,-1cm)}]{{Projection (line 8)}/projection,{$\beta$ step (line 9)}/betastep}
	\node at (1.2cm,1.8cm) {(a)};
	\node at (6cm,1.8cm) {(b)};
\end{tikzpicture}
\end{center}
\caption{Computation times for (a) single-core and (b) parallel implementations. Unnumbered slices contribute with 1\%.}
\label{fig:cycles}
\end{figure*}
A well-conditioned Hessian can also be obtained from choosing appropriate $\mathbf{Q}$ and $\mathbf{R}$. For $N\!=\!1$, the Hessian is $\mathbf{J}=\mathbf{B}^\Tr \mathbf{P} \mathbf{B}+\mathbf{R}$ and the analysis can be greatly simplified by transforming system~\eqref{eq:ssshort} into modal space, i.e. by approximating $\mathbf{A}\approx aI$, $\mathbf{B}\approx bI$ and using $\mathbf{C}=\mathbf{U}\mathbf{\Sigma}\trans{\mathbf{V}}$ to diagonalize~\eqref{eq:ssshort}. In modal space, the matrix $\mathbf{\hat{P}}=\diag(\hat{p}_1,\dots,\hat{p}_{n_u})\eqdef\trans{\mathbf{V}}\mathbf{P}\mathbf{V}$ is diagonal and the DARE~\eqref{eq:dare} is solved by
\begin{align}\label{eq:modaldare}
\hat{p}_i = \frac{1}{2b^2}\left(-\xi_i+\sqrt{\xi_i^2+4b^2 \hat{q}_i\hat{r}_i}\right),
\end{align}
where $\xi_i =\hat{r}_i-a^2\hat{r}_i-b^2\hat{q}_i$ and $\hat{q}_i$ and $\hat{r}_i$ are the diagonal elements of $\mathbf{\hat{Q}}\eqdef\trans{\mathbf{\Sigma}}\mathbf{\Sigma}$ and $\mathbf{\hat{R}}\eqdef\diag(\hat{r}_1,\dots,\hat{r}_{n_u})$. For fixed $\hat{r}_i$, the low-order modes (large $\hat{q}_i$) have a large cost (large $\hat{p}_i$), which in turn yields a large LQR control gain $\hat{k}_i=ab\hat{p}_i/(\hat{r}_i+b^2\hat{p}_i)$~\cite[Ch. 9.2]{SKOGESTADMULTI}. The DLS-I \emph{internal model controller} (IMC) is closely related to LQR~\cite{MORARIIMC1} and computes the open-loop gains as $\invbr{\trans{\mathbf{\Sigma}}\mathbf{\Sigma}+\lambda I}\trans{\mathbf{\Sigma}}$, where $\lambda>0$ is a regularization parameter. The modal input weights $\hat{r}_i$ can be chosen such that the LQR controller gain matches the IMC open-loop gain for each mode $i$, which is depicted in Fig.~\ref{fig:weights} (a) and (b) (IMC, red). For IMC with $\lambda = 0$, the open-loop gain is proportional to $\inv{\mathbf{\Sigma}}$, so the higher order modes must be detuned, whereas for LQR, the open-loop gain is ``proportional'' to $\mathbf{\Sigma}$, so the low order modes must be detuned. However, this choice of input weights does not decrease the condition number of the Hessian. For $N\!=\!1$, the condition number of the resulting Hessian is $7,485$ and $11,616$ for $N\!=\!2$.

A simple way to significantly decrease the condition number of the Hessian is to choose $\mathbf{\hat{R}}=I$ and to limit the diagonal elements of $\mathbf{\hat{Q}}$ to a minimum and maximum value, which is depicted in Fig.~\ref{fig:weights}.a (saturated, blue). This approach has also been chosen in Fig.~\ref{fig:ibm}, where it can be seen that there is no decrease in controller performance compared to IMC. For $N\!=\!1$, the condition number of the resulting Hessian is $21$, while it is $31$ for $N\!=\!2$. Note that the state weighting matrix in the original space can be recovered by setting $\mathbf{Q}=\mathbf{V}\mathbf{\hat{Q}}\trans{\mathbf{V}}$.

The required number of iterations for Alg.~\ref{alg:fgm} is illustrated in Fig.~\ref{fig:weights}.b, which shows the number of iterations averaged over 10,000 MPC problem instances. For each instance we count the number of iterations required for the algorithm's iterates to satisfy $\infnorm{\mathbf{p}_{i+1}-\mathbf{p}_{i}}<\epsilon$ and $\infnorm{\mathbf{p}_{i+1}-\mathbf{p}_{i}}<\epsilon\infnorm{\mathbf{p}_{i}}$ with $\epsilon=10^{\sm 3}$. As expected from the upper bound~\eqref{eq:iterbound}, significantly more iterations are required when the condition number is large.
\section{Implementation}\label{sec:implementation}
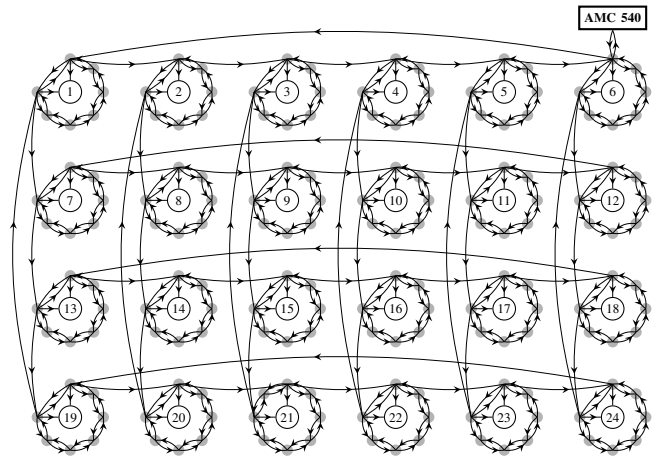
\begin{figure}
\begin{center}
\begin{tikzpicture}[%
node distance=2mm,
bpm/.style={fill=colorlightgray,minimum width=1.5mm,minimum height=1.5mm,inner sep=0,circle},%
cor/.style={draw,circle,minimum width=3mm,minimum height=3mm,inner sep=0,text centered,font=\tiny},%
amc/.style={draw,thick,minimum width=3mm,minimum height=3mm,inner sep=2,text centered,font=\tiny\bfseries}]

\node[cor] 						(c1) 	{1};
\node[bpm,left=of c1] 			(b1l) 	{};
\node[bpm,below=of c1] 			(b1b) 	{};
\node[bpm,below right=of c1] 	(b1br) 	{};
\node[bpm,below left=of c1] 	(b1bl) 	{};
\node[bpm,above=of c1] 			(b1t) 	{};
\node[bpm,above right=of c1] 	(b1tr) 	{};
\node[bpm,right=of c1] 			(b1r) 	{};

\node[bpm,below=4mm of b1b]		(b7t) 	{};
\node[cor,below=of b7t]			(c7) 	{7};
\node[bpm,left=of c7] 			(b7l) 	{};
\node[bpm,below=of c7] 			(b7b) 	{};
\node[bpm,below right=of c7] 	(b7br) 	{};
\node[bpm,below left=of c7] 	(b7bl) 	{};
\node[bpm,above right=of c7] 	(b7tr) 	{};
\node[bpm,right=of c7] 			(b7r) 	{};

\node[bpm,below=4mm of b7b]		(b13t) 		{};
\node[cor,below=of b13t]		(c13) 		{13};
\node[bpm,left=of c13] 			(b13l) 		{};
\node[bpm,below=of c13]			(b13b) 		{};
\node[bpm,below right=of c13] 	(b13br) 	{};
\node[bpm,below left=of c13] 	(b13bl) 	{};
\node[bpm,above right=of c13] 	(b13tr) 	{};
\node[bpm,right=of c13]			(b13r) 		{};

\node[bpm,below=4mm of b13b]	(b19t) 		{};
\node[cor,below=of b19t]		(c19) 		{19};
\node[bpm,left=of c19] 			(b19l) 		{};
\node[bpm,below=of c19]			(b19b) 		{};
\node[bpm,below right=of c19] 	(b19br) 	{};
\node[bpm,below left=of c19] 	(b19bl) 	{};
\node[bpm,above right=of c19] 	(b19tr) 	{};
\node[bpm,right=of c19]			(b19r) 		{};

\node[bpm,right=4mm of b1r]		(b2l) 	{};
\node[cor,right=of b2l]			(c2) 	{2};
\node[bpm,below=of c2] 			(b2b) 	{};
\node[bpm,below right=of c2] 	(b2br) 	{};
\node[bpm,below left=of c2] 	(b2bl) 	{};
\node[bpm,above=of c2] 			(b2t) 	{};
\node[bpm,above right=of c2] 	(b2tr) 	{};
\node[bpm,right=of c2] 			(b2r) 	{};

\node[bpm,below=4mm of b2b]		(b8t) 	{};
\node[cor,below=of b8t]			(c8) 	{8};
\node[bpm,left=of c8] 			(b8l) 	{};
\node[bpm,below=of c8] 			(b8b) 	{};
\node[bpm,below right=of c8] 	(b8br) 	{};
\node[bpm,below left=of c8] 	(b8bl) 	{};
\node[bpm,above right=of c8] 	(b8tr) 	{};
\node[bpm,right=of c8] 			(b8r) 	{};

\node[bpm,below=4mm of b8b]		(b14t) 		{};
\node[cor,below=of b14t]		(c14) 		{14};
\node[bpm,left=of c14] 			(b14l) 		{};
\node[bpm,below=of c14]			(b14b) 		{};
\node[bpm,below right=of c14] 	(b14br) 	{};
\node[bpm,below left=of c14] 	(b14bl) 	{};
\node[bpm,above right=of c14] 	(b14tr) 	{};
\node[bpm,right=of c14]			(b14r) 		{};

\node[bpm,below=4mm of b14b]	(b20t) 		{};
\node[cor,below=of b20t]		(c20) 		{20};
\node[bpm,left=of c20] 			(b20l) 		{};
\node[bpm,below=of c20]			(b20b) 		{};
\node[bpm,below right=of c20] 	(b20br) 	{};
\node[bpm,below left=of c20] 	(b20bl) 	{};
\node[bpm,above right=of c20] 	(b20tr) 	{};
\node[bpm,right=of c20]			(b20r) 		{};

\node[bpm,right=4mm of b2r]		(b3l) 	{};
\node[cor,right=of b3l]			(c3) 	{3};
\node[bpm,below=of c3] 			(b3b) 	{};
\node[bpm,below right=of c3] 	(b3br) 	{};
\node[bpm,below left=of c3] 	(b3bl) 	{};
\node[bpm,above=of c3] 			(b3t) 	{};
\node[bpm,above right=of c3] 	(b3tr) 	{};
\node[bpm,right=of c3] 			(b3r) 	{};

\node[bpm,below=4mm of b3b]		(b9t) 	{};
\node[cor,below=of b9t]			(c9) 	{9};
\node[bpm,left=of c9] 			(b9l) 	{};
\node[bpm,below=of c9] 			(b9b) 	{};
\node[bpm,below right=of c9] 	(b9br) 	{};
\node[bpm,below left=of c9] 	(b9bl) 	{};
\node[bpm,above right=of c9] 	(b9tr) 	{};
\node[bpm,right=of c9] 			(b9r) 	{};

\node[bpm,below=4mm of b9b]		(b15t) 		{};
\node[cor,below=of b15t]		(c15) 		{15};
\node[bpm,left=of c15] 			(b15l) 		{};
\node[bpm,below=of c15]			(b15b) 		{};
\node[bpm,below right=of c15] 	(b15br) 	{};
\node[bpm,below left=of c15] 	(b15bl) 	{};
\node[bpm,above right=of c15] 	(b15tr) 	{};
\node[bpm,right=of c15]			(b15r) 		{};

\node[bpm,below=4mm of b15b]	(b21t) 		{};
\node[cor,below=of b21t]		(c21) 		{21};
\node[bpm,left=of c21] 			(b21l) 		{};
\node[bpm,below=of c21]			(b21b) 		{};
\node[bpm,below right=of c21] 	(b21br) 	{};
\node[bpm,below left=of c21] 	(b21bl) 	{};
\node[bpm,above right=of c21] 	(b21tr) 	{};
\node[bpm,right=of c21]			(b21r) 		{};
\node[bpm,above left=of c21] 	(b21tl) 	{};

\node[bpm,right=4mm of b3r]		(b4l) 	{};
\node[cor,right=of b4l]			(c4) 	{4};
\node[bpm,below=of c4] 			(b4b) 	{};
\node[bpm,below right=of c4] 	(b4br) 	{};
\node[bpm,below left=of c4] 	(b4bl) 	{};
\node[bpm,above=of c4] 			(b4t) 	{};
\node[bpm,above right=of c4] 	(b4tr) 	{};
\node[bpm,right=of c4] 			(b4r) 	{};

\node[bpm,below=4mm of b4b]		(b10t) 	{};
\node[cor,below=of b10t]		(c10) 	{10};
\node[bpm,left=of c10] 			(b10l) 	{};
\node[bpm,below=of c10]			(b10b) 	{};
\node[bpm,below right=of c10] 	(b10br)	{};
\node[bpm,below left=of c10] 	(b10bl)	{};
\node[bpm,above right=of c10] 	(b10tr)	{};
\node[bpm,right=of c10]			(b10r) 	{};

\node[bpm,below=4mm of b10b]	(b16t) 		{};
\node[cor,below=of b16t]		(c16) 		{16};
\node[bpm,left=of c16] 			(b16l) 		{};
\node[bpm,below=of c16]			(b16b) 		{};
\node[bpm,below right=of c16] 	(b16br) 	{};
\node[bpm,below left=of c16] 	(b16bl) 	{};
\node[bpm,above right=of c16] 	(b16tr) 	{};
\node[bpm,right=of c16]			(b16r) 		{};

\node[bpm,below=4mm of b16b]	(b22t) 		{};
\node[cor,below=of b22t]		(c22) 		{22};
\node[bpm,left=of c22] 			(b22l) 		{};
\node[bpm,below=of c22]			(b22b) 		{};
\node[bpm,below right=of c22] 	(b22br) 	{};
\node[bpm,below left=of c22] 	(b22bl) 	{};
\node[bpm,above right=of c22] 	(b22tr) 	{};
\node[bpm,right=of c22]			(b22r) 		{};

\node[bpm,right=4mm of b4r]		(b5l) 	{};
\node[cor,right=of b5l]			(c5) 	{5};
\node[bpm,below=of c5] 			(b5b) 	{};
\node[bpm,below right=of c5] 	(b5br) 	{};
\node[bpm,below left=of c5] 	(b5bl) 	{};
\node[bpm,above=of c5] 			(b5t) 	{};
\node[bpm,above right=of c5] 	(b5tr) 	{};
\node[bpm,right=of c5] 			(b5r) 	{};

\node[bpm,below=4mm of b5b]		(b11t) 	{};
\node[cor,below=of b11t]		(c11) 	{11};
\node[bpm,left=of c11] 			(b11l) 	{};
\node[bpm,below=of c11]			(b11b) 	{};
\node[bpm,below right=of c11] 	(b11br)	{};
\node[bpm,below left=of c11] 	(b11bl)	{};
\node[bpm,above right=of c11] 	(b11tr)	{};
\node[bpm,right=of c11]			(b11r) 	{};

\node[bpm,below=4mm of b11b]	(b17t) 		{};
\node[cor,below=of b17t]		(c17) 		{17};
\node[bpm,left=of c17] 			(b17l) 		{};
\node[bpm,below=of c17]			(b17b) 		{};
\node[bpm,below right=of c17] 	(b17br) 	{};
\node[bpm,below left=of c17] 	(b17bl) 	{};
\node[bpm,above right=of c17] 	(b17tr) 	{};
\node[bpm,right=of c17]			(b17r) 		{};

\node[bpm,below=4mm of b17b]	(b23t) 		{};
\node[cor,below=of b23t]		(c23) 		{23};
\node[bpm,left=of c23] 			(b23l) 		{};
\node[bpm,below=of c23]			(b23b) 		{};
\node[bpm,below right=of c23] 	(b23br) 	{};
\node[bpm,below left=of c23] 	(b23bl) 	{};
\node[bpm,above right=of c23] 	(b23tr) 	{};
\node[bpm,right=of c23]			(b23r) 		{};

\node[bpm,right=4mm of b5r]		(b6l) 	{};
\node[cor,right=of b6l]			(c6) 	{6};
\node[bpm,below=of c6] 			(b6b) 	{};
\node[bpm,below right=of c6] 	(b6br) 	{};
\node[bpm,below left=of c6] 	(b6bl) 	{};
\node[bpm,above=of c6] 			(b6t) 	{};
\node[bpm,above right=of c6] 	(b6tr) 	{};
\node[bpm,right=of c6] 			(b6r) 	{};

\node[bpm,below=4mm of b6b]		(b12t) 	{};
\node[cor,below=of b12t]		(c12) 	{12};
\node[bpm,left=of c12] 			(b12l) 	{};
\node[bpm,below=of c12]			(b12b) 	{};
\node[bpm,below right=of c12] 	(b12br)	{};
\node[bpm,below left=of c12] 	(b12bl)	{};
\node[bpm,above right=of c12] 	(b12tr)	{};
\node[bpm,right=of c12]			(b12r) 	{};

\node[bpm,below=4mm of b12b]	(b18t) 		{};
\node[cor,below=of b18t]		(c18) 		{18};
\node[bpm,left=of c18] 			(b18l) 		{};
\node[bpm,below=of c18]			(b18b) 		{};
\node[bpm,below right=of c18] 	(b18br) 	{};
\node[bpm,below left=of c18] 	(b18bl) 	{};
\node[bpm,above right=of c18] 	(b18tr) 	{};
\node[bpm,right=of c18]			(b18r) 		{};

\node[bpm,below=4mm of b18b]	(b24t) 		{};
\node[cor,below=of b24t]		(c24) 		{24};
\node[bpm,left=of c24] 			(b24l) 		{};
\node[bpm,below=of c24]			(b24b) 		{};
\node[bpm,below right=of c24] 	(b24br) 	{};
\node[bpm,below left=of c24] 	(b24bl) 	{};
\node[bpm,above right=of c24] 	(b24tr) 	{};
\node[bpm,right=of c24]			(b24r) 		{};

\begin{scope}[very thin,decoration={
    markings,
    mark=at position 0.8 with {\arrow{stealth}}}
    ]
\foreach \x in {1,...,24} {
\draw[postaction={decorate}] (b\x l.center)--(c\x.west);
\draw[postaction={decorate}] (b\x t.center)--(c\x.north);
\draw[postaction={decorate}] (b\x l.center)to[bend right](b\x bl.center);
\draw[postaction={decorate}] (b\x bl.center)to[bend right](b\x l.center);
\draw[postaction={decorate}] (b\x bl.center)to[bend right](b\x b.center);
\draw[postaction={decorate}] (b\x b.center)to[bend right](b\x bl.center);
\draw[postaction={decorate}] (b\x b.center)to[bend right](b\x br.center);
\draw[postaction={decorate}] (b\x br.center)to[bend right](b\x b.center);
\draw[postaction={decorate}] (b\x br.center)to[bend right](b\x r.center);
\draw[postaction={decorate}] (b\x r.center)to[bend right](b\x br.center);
\draw[postaction={decorate}] (b\x r.center)to[bend right](b\x tr.center);
\draw[postaction={decorate}] (b\x tr.center)to[bend right](b\x r.center);
\draw[postaction={decorate}] (b\x tr.center)to[bend right](b\x t.center);
\draw[postaction={decorate}] (b\x t.center)to[bend right](b\x tr.center);
};

\draw[postaction={decorate}] (b21l.center)to[bend right](b21tl.center);
\draw[postaction={decorate}] (b21tl.center)to[bend right](b21l.center);
\draw[postaction={decorate}] (b21tl.center)to[bend right](b21t.center);
\draw[postaction={decorate}] (b21t.center)to[bend right](b21tl.center);
\end{scope}

\begin{scope}[very thin,decoration={
    markings,
    mark=at position 0.5 with {\arrow{stealth}}}
    ]
\foreach \x in {1,...,24} {
\ifnum\x = 21

\else
\draw[postaction={decorate}] (b\x l.center)to[bend right=15](b\x t.center);
\draw[postaction={decorate}] (b\x t.center)to[bend right=15](b\x l.center);
\fi
};
\end{scope}

\begin{scope}[very thin,decoration={
    markings,
    mark=at position 0.6 with {\arrow{stealth}}}
    ]
\foreach \y in {0,...,3} {
\foreach \x in {1,...,5} {
\pgfmathtruncatemacro{\curr}{\x + \y*6}
\pgfmathtruncatemacro{\next}{\curr + 1}
\draw[postaction={decorate}] (b\curr t.center)to[bend right=10](b\next t.center);
};
};
\end{scope}

\begin{scope}[very thin,decoration={
    markings,
    mark=at position 0.6 with {\arrow{stealth}}}
    ]
\foreach \y in {0,...,5} {
\foreach \x in {0,...,2} {
\pgfmathtruncatemacro{\curr}{1+\x*6 + \y}
\pgfmathtruncatemacro{\next}{\curr + 6}
\draw[postaction={decorate}] (b\curr l.center)to[bend right=10](b\next l.center);
};
};
\end{scope}

\begin{scope}[very thin,decoration={
    markings,
    mark=at position 0.6 with {\arrow{stealth}}}
    ]
\foreach \x in {19,...,24} {
\pgfmathtruncatemacro{\curr}{\x}
\pgfmathtruncatemacro{\next}{\curr - 18}
\draw[postaction={decorate}] (b\curr l.center)to[bend left=15](b\next l.center);
};
\end{scope}

\begin{scope}[very thin,decoration={
    markings,
    mark=at position 0.55 with {\arrow{stealth}}}
    ]
\foreach \x in {6,12,18,24} {
\pgfmathtruncatemacro{\curr}{\x}
\pgfmathtruncatemacro{\next}{\curr - 5}
\draw[postaction={decorate}] (b\curr t.center)to[bend right=10](b\next t.center);
};
\end{scope}

\node[amc, above=3mm of b6t] (a) {AMC 540};
\begin{scope}[very thin,decoration={
    markings,
    mark=at position 0.6 with {\arrow{stealth}}}
    ]
\draw[postaction={decorate}] (b6t.center)to[bend right=20](a.south);
\draw[postaction={decorate}] (a.south)to[bend right=20](b6t.center);
\end{scope}
\end{tikzpicture}
\end{center}
\caption{Diamond-I communication network topology.}\label{fig:FOFB}
\end{figure}

 DLS-I has implemented the network topology shown in Fig.~\ref{fig:FOFB} for transmitting the BPM measurements across the storage ring. At each time instant, the BPMs (gray dots) inject new measurements into the network, which are synchronized and forwarded to each of the 24 nodes that compute the control inputs for the neighboring corrector magnets. DLS-II will considerably simplify the topology of Fig.~\ref{fig:FOFB} and implement a centralized network, where the BPM signals from each cell will be sent to one central computing node. For testing our algorithm on DLS-I, we connect the new hardware to the communication network as illustrated in Fig.~\ref{fig:FOFB}. The computed control signals will then be ``disguised'' as BPM signals again and each of the 24 distributed nodes will select the corresponding signal to pass to the neighboring magnets.
The new central computing node is a VadaTech AMC540 board~\cite{AMC540} that embeds a Xilinx Virtex-7 FPGA and two Texas Instruments (TI) C6678 digital signal processors (DSPs)~\cite{TIC6678} with 8 cores each. For our tests, the control algorithm will be implemented on the DSPs, which are more flexible to program, while the FPGA will be responsible for signal routing. A PCIe link is used to transfer BPM and control input data between the FPGA and the DSPs, which takes roughly $5~\mu$s ($6.6$~Gbps) when executed by the direct memory access (DMA) engine of the DSP. The DSPs are clocked at 1.4~GHz and the sampling frequency of $10$~kHz allows for 140,000 processor cycles ($100$~$\mu$s). One core of each DSP is used to communicate with the control room through a gigabit ethernet link. The control problems for the vertical and horizontal beam directions are independent and one DSP is used for each direction.
\subsection{Single-Core Implementation}
The TI C6678 is a floating point processor with single-instructions multiple-data (SIMD) capabilities that can be programmed in C. It has two levels of core-local memory (L1, 32~kB and L2, 512~kB) and a third level of shared memory (L3, 4~MB) with the L1 memory being configured as cache. Accessing the L2 memory is twice as fast as accessing the L3 memory~\cite{TIC6678}.

For the gradient step of Alg.~\ref{alg:fgm}, we have implemented a highly optimized routine that exploits the core architecture and uses SIMDs. Analogous to standard row-major matrix-vector multiplication, the routine implements two nested for-loops, where the first loop iterates over rows and the second over columns. To minimize memory transactions and maximize the use of SIMD, the inner loop computes 8 rows and 4 columns at once. To maximize the efficiency of the cache, the arrays are aligned to cache line boundaries, zero-padded to multiples of $4$ or $8$ floats and rearranged such that the unrolled rows are contiguous in memory.

For $N\!=\!1$, the algorithm can be implemented as shown in Alg.~\ref{alg:fgm} and all the problem data, such as the Hessian $\mathbf{J}$, can be saved in L2 memory. For $N\!=\!2$, the Hessian uses almost the whole L2 memory, so some data must be moved to the slower L3 memory. The cache efficiency for the projection can be increased by permuting the data using a perfect shuffle, so that the inputs for magnet $i$ and horizon stages $0$ and $1$ are contiguous in memory. The computational complexity of the gradient step could be reduced by considering the sparsity patterns in the definition of the Hessian~\eqref{eq:Jdef}, i.e. by separating the multiplications by $\mathbf{G}$ and $(I_N\otimes \mathbf{Q})\oplus \mathbf{P}$.

The single-core performance with $I_\text{max}=20$ and horizon $N\!=\!\lbrace 1,2\rbrace$ is shown in Fig.~\ref{fig:cycles}.a. It requires $543$~$\mu$s for $N\!=\!1$ and $3550$~$\mu$s for $N\!=\!2$ to compute the control inputs, which is more than the desired $100$~$\mu$s. The most expensive operation is the gradient step, which takes $357$~$\mu$s for $N\!=\!1$ and $3000$~$\mu$s for $N\!=\!2$. As the algorithm is dominated by the gradient step, one would expect the computation time to quadruple when doubling the problem size. However, transferring problem data that lies in the L3 memory and additional cache inefficiencies incur substantial overheads. The single core performance could certainly be increased by configuring the L2 memory as cache, but our parallel implementation uses the L2 memory for saving core-local data and this approach was not pursued further.
\subsection{Parallelization}\label{sec:par}
All steps of Alg.~\ref{alg:fgm} can be parallelized using a standard manager-worker framework, but variable dependencies require core communication and cache operations that are denoted by circled arrows. The same executable is used for all cores and the code is branched off based on the core ID. Note that the observer operations $\mathbf{\hat{y}}_k\eqdef\mathbf{C}\mathbf{\hat{z}}_k^\mu$ and $\mathbf{L}(\mathbf{y}_k-\mathbf{\hat{y}}_k-\mathbf{\hat{d}}_k)$ are computed separately and require two synchronization steps.

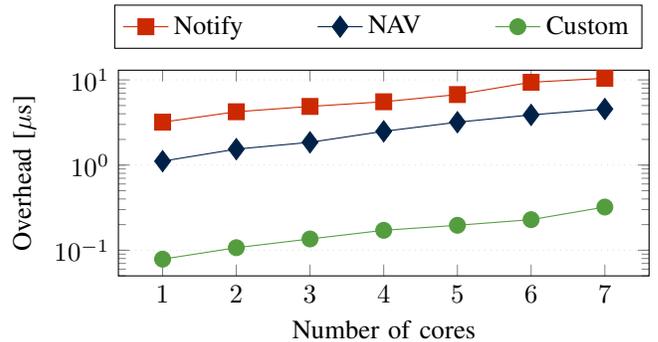
\begin{figure}
{\hspace{1.49cm}\ref{ovrlegend}\vspace{0.15cm}}
\begin{center}
\begin{tikzpicture}
\begin{semilogyaxis}[xlabel=Number of cores,
xtick={1,2,3,4,5,6,7},
ylabel={Overhead [$\mu$s]},
legend pos=north west,
name=plot1,
major grid style={line width=.2pt,draw=black},
width=\linewidth,
height=0.5\linewidth,
legend columns = 3,
legend cell align={left},
legend style={/tikz/every even column/.append style={column sep=0.91cm}},
legend to name=ovrlegend,
ymin=5*10^(-2), ymax=13,
ymajorgrids,
major grid style={line width=.2pt,draw=gray!50,dotted},
]
\addplot [color=colorred, mark=square*, mark size=3] table [x=a, y=b, col sep=comma] {ipc_times_mus.csv};
\addlegendentry{Notify}
\addplot [color=colorblue, mark=diamond*, mark size=4] table [x=a, y=c, col sep=comma] {ipc_times_mus.csv};
\addlegendentry{NAV}
\addplot [color=colorgreen, mark=*, mark size=3] table [x=a, y=d, col sep=comma] {ipc_times_mus.csv};
\addlegendentry{Custom}
\end{semilogyaxis}
\end{tikzpicture}\end{center}
\caption{Interprocessor communication overhead.}\label{fig:ovr}
\end{figure}

For the problem size of the MPC problem~\eqref{eq:qpshort}, the cost of parallelization is not negligible. Fig.~\ref{fig:ovr} shows the overhead introduced by interprocessor communication measured by the elapsed time between a manager request and the acknowledgement of $n_w$ worker cores without worker payload. Three different implementations are compared: The TI Notify scheme, which is a library provided by TI and used by the TI open multi-processing (openMP) toolbox, the TI multicore navigator (NAV), which is implemented through a separate on-chip processor, and our custom interrupt-free implementation. The TI notification schemes are flexible, but introduce a considerable delay. Note that with 20 synchronization points, the TI Notify scheme alone would introduce $200$~$\mu$s of overhead. For our custom approach, we chose to implement a simpler scheme using integer flags that are saved in the L3 memory. For further speed-up, the L1 cache is by-passed by creating a non-cacheable virtual memory section. In practice, at each communication step it is also required to invalidate or write-back the cache, which can be manually triggered using TI's chip support library.

Alg.~\ref{alg:fgm} is sliced into $6\times 32$ row-blocks with $192$ columns each and deployed on 6 worker cores and 1 manager core. The length of the slices must be a multiple of the cache line size ($64$~B) and using 7 worker cores would not yield any speed up. The master core coordinates the various steps of Alg.~\ref{alg:fgm}, communicates with the adjacent FPGA and triggers the DMA. A breakdown of the computation time of Alg.~\ref{alg:fgm} with $I_\text{max}=20$ is shown in Fig.~\ref{fig:cycles}.b. For $N\!=\!1$, the algorithm uses $69$~$\mu$s, which is well below the allowed $100$~$\mu$s, but for $N\!=\!2$, the computation time of $272$~$\mu$s is far above the time limit. 

Comparing Fig.~\ref{fig:cycles}.a and b, the parallelization reduces the computation time by a factor between about $8$ and $13$. In theory, one would expect the computation time to be reduced by a factor smaller than $n_w$ when deployed onto $n_w$ worker cores. We suspect that this discrepancy is due to memory and cache bandwidth limitations on the single core implementation.
\section{Conclusion}
In this feasibility study, we have focused on the practical issues of implementing MPC for the DLS-I electron beam stabilization problem. To obtain an implementation that runs at the desired speed, we tailored the MPC algorithm to the application. Firstly, we avoided removing the time delay by augmenting the system with additional states and designed an observer for the delayed states instead. The delayed measurement updates were then projected into the future, which exploited the diagonal structure of the state-space system. Secondly, because standard preconditioning techniques with diagonal preconditioning matrices were unable to reduce the condition number of the Hessian, we used the modal decomposition to choose appropriate state- and input-weighting matrices that led to a Hessian with a small condition number. Finally, we showed that standard parallelization toolboxes, such as openMP, introduce overheads that would prohibit the algorithm from running at the desired speed, and we therefore implemented a customized core-synchronization framework. Our investigation showed that MPC is applicable to the electron beam stabilization problem, but requires investment of significant effort into the theoretical and practical implementation as well as paying particular attention to details, such as overheads introduced by the CQP initialization or parallelization, which are often neglected in theoretical investigations. Our practical tests also showed that assumptions on computational complexities can be inaccurate, e.g. doubling the CQP problem size does not necessarily result in quadruple computation time nor does parallelizing the algorithm on $n_w$ cores increase the computation speed by a factor of $n_w$.

In anticipation of our tests, we demonstrated the feasibility for the DLS-I storage ring, but we have not considered a number of additional changes that DLS-II will introduce. The number of actuators will be increased from $173$ to $396$ for Diamond-II, which will significantly increase the computational complexity of the algorithm and further slow down the controller. However, in contrast to the current system the DLS-II system will have a block-circulant and centrosymmetric symmetry, which can be exploited to increase the computational speed of the controller by a factor of $10$~\cite{SYNCSYM}.

For DLS-I, all corrector magnets are actuated at $10$~kHz, whereas at Diamond-II, the $144$ fast actuators will be actuated at $100$~kHz and the slow actuators at $1$~kHz. This would give rise to another MPC scheme in which the control inputs for the slow actuators are computed every $100$ time step and the control inputs for the fast actuators are computed every other time step. For such an MPC scheme, the closed-loop stability would need to be assessed separately.

A communication controller in the DLS-I storage ring manages the communication between computing nodes and BPMs. At DLS-II, the BPM measurements will be sent to one central node and not all measurements will be synchronized. In this paper, we designed an observer that receives measurements that have the same time delay and projects the measurement update to the current state. If the measurement have different delays, this could be considered in the observer.

All our simulations used measurement data from DLS-I and it is expected that the power spectrum of the DLS-II disturbances will change. A disturbance model was used to compute the feedforward setpoint, and it was assumed that the disturbances are independent and identically distributed. For DLS-II, the disturbances might be correlated, in which case a different disturbance model could be used. Considering correlated disturbances could increase the performance of the controller in terms of disturbance attenuation.

{\bibliographystyle{IEEEtran}
\small
\bibliography{IEEEabbrv,master_bib_abbrev}}
\end{document}